\documentclass{rstransa}

\usepackage{amssymb,amsmath,color,psfrag,hyperref}
\usepackage{environ}
\usepackage{wasysym,empheq}
\usepackage{comment}
\usepackage{notoccite}

\jname{rsta}
\Journal{Phil. Trans. R. Soc}

\newcommand{\la}{\langle}
\newcommand{\ra}{\rangle}
\newcommand{\lb}{\left|}
\newcommand{\rb}{\right|}
\newcommand{\lnb}{\left(}
\newcommand{\rnb}{\right)}
\newcommand{\lsqb}{\left[}
\newcommand{\rsqb}{\right]}

\begin{document}

\title{Maxwell electromagnetism as an emergent phenomenon in condensed matter}

\author{J. Rehn$^{1}$, R. Moessner$^{1}$}
\address{$^{1}$Max-Planck-Institut f\"ur Physik komplexer Systeme, 01187 Dresden, Germany}

%%%% Subject entries to be placed here %%%%
%\subject{xxxxx, xxxxx, xxxx}

%%%% Keyword entries to be placed here %%%%
%\keywords{xxxx, xxxx, xxxx}

%%%% Insert corresponding author and its email address}
%\corres{Insert corresponding author name\\
%\email{xxx@xxxx.xx.xx}}

\begin{abstract}
  The formulation of a complete theory of classical
  electromagnetism by Maxwell is one of the milestones of science.
  The capacity of many-body systems to provide emergent mini-universes
  with vacua quite distinct from the one we inhabit was only
  recognised much later. Here, we provide an account of how simple
  systems of localised spins manage to emulate Maxwell
  electromagnetism in their low-energy behaviour. They are much
  less constrained by symmetry considerations than the relativistically
  invariant electromagnetic vacuum, as their substrate
  provides a non-relativistic background with even translational
  invariance broken. They can exhibit rich behaviour not encountered in conventional electromagnetism.
  This includes the existence of magnetic monopole excitations arising 
  from fractionalisation of magnetic dipoles; as well as the
  capacity of disorder, by generating defects on the lattice scale,
  to produce novel physics, as exemplified by topological spin glassiness or
  random Coulomb magnetism.
\end{abstract}

%%%%%%%%%%%%%%% End of first page %%%%%%%%%%%%%%%%%%%%%

\maketitle

%==============================================================================%
\section{Introduction}
The tremendous importance of Maxwell's equations is hard to state,
let alone overstate. They underpin a large fraction of our sensory
perceptions, those related to light, are at the root of literally
innumerable applications of electromagnetism, and provide the rung
above gravity in the ladder of fundamental interactions, which
climbing up further has brought forth  the spectacular successes
of high energy physics, in particular the standard model, and led
us to string theory.

It was only in the twentieth century that it was fully recognised
that much is to be gained from moving in the opposite direction
in energy. While particle accelerators got larger and more
powerful in order to push back the frontier of high energy
experiments, it was by cooling that many phenomena in solid
state physics were discovered which account for much of the
interest in that field. The liquefaction of helium, and the
resulting discoveries of superfluidity and superconductivity
a century ago, are representatives for a string of many
important developments.

We are now used to, as a matter of course that hardly requires
comment, the fact that low-energy degrees of freedom of a
system can differ completely from those of its high-energy
constituents. This phenomenon is now known under the name
emergence, and was formulated most crisply in Anderson's
manifesto entitled `More is different'~\cite{anderson1972more}.
This independence of degrees of freedom across scales is
crisply noticed in the case of crystals, whose low-energy
excitations--the phononic lattice vibrations--exhibit bosonic
quantum statistics, regardless of the fermionic or bosonic
nature of the atoms making up the crystal.

The capacity of phase transitions -- such as the translational
symmetry breaking leading to the phonons in the above example -- to
generate various types of emergent excitations being well established,
it perhaps comes as a surprise that there are not all that many
experimental instances where these degrees of freedom are not a
fermionic or bosonic field but instead gauge fields. This
experimental scarcity is not due to theoretical neglect, there
being no shortage of models where their existence is, if not
established, at least fervently desired.

This article is devoted to an account of the emergence of
Maxwell electromagnetism in a class of magnetic systems known
as highly frustrated magnets. In these systems, competing
interactions yield a large number of nearly degenerate
low-energy states; fluctuations between these lead to new
collective behaviour which in some cases turns out to be
naturally captured in a description based on an emergent
gauge field. 

In the following, in Sec.~\ref{sec:emgauge}, we account for
how this comes about in some detail; and then focus on
obvervable consequences, placing some emphasis on aspects
of the emergent physics which are not known to have a
counterpart in conventional electromagnetism. These include
the existence of magnetically charged quasiparticles known
as emergent magnetic monopoles. Also, a selection of
disorder-induced phenomena is covered in Sec.~\ref{sec:disSL},
where the interplay of lattice scale defects to the `ether'
of emergent electromagnetism and its long-wavelength Coulomb
field is discussed: disorder can nucleate gauge-charged
defects, the interactions between which lead to new collective
disorder physics.
We conclude with an outlook in Sec.~\ref{sec:concl}.

The material in this article has been selected rather
idiosyncratically in order to provide a flavour of this field to the
interested outsider; it is neither comprehensive in its choice of
topics, nor a complete historical account. For a more detailed
pedagogical introduction, we mention review articles on geometrically
frustrated magnets~\cite{chalker2009geometrical} and their Coulomb
phase~\cite{henley2010coulomb} and spin
ice~\cite{castelnovo2012spin}.

\section{Emergent gauge fields}
\label{sec:emgauge}

Antiferromagnets present an impressive variety
of possible ground state configurations, crucially depending
on the geometry of the lattice where the spins reside.
This is related to the fact that interactions are
frustrated once the lattice contains odd length loops
and the ground state configurations become highly degenerate.
This high degeneracy can usually be traced back to
an underconstraint in the system of equations found by
minimising the Hamiltonian \cite{moessner1998properties}. This idea,
amusingly enough, can also be traced back to an idea of 
Maxwell's \cite{MaxwellConstraint}.

The intricate structure of correlations within the degenerate set of 
ground states will give the excitations special properties.

On some systems the ground state
correlations can be interpreted as arising from constraints
defining emerging conservation laws, which are then
resolved by a gauge field.
This mapping from the original spin variables to the
emergent gauge field leads not only to an understanding
of the correlations within the set of ground state
configurations, but also allows prediction of the behavior
of the local excitations.

\begin{figure}
    \centering
	\includegraphics[width=1.2\dimexpr11\textwidth/16\relax]{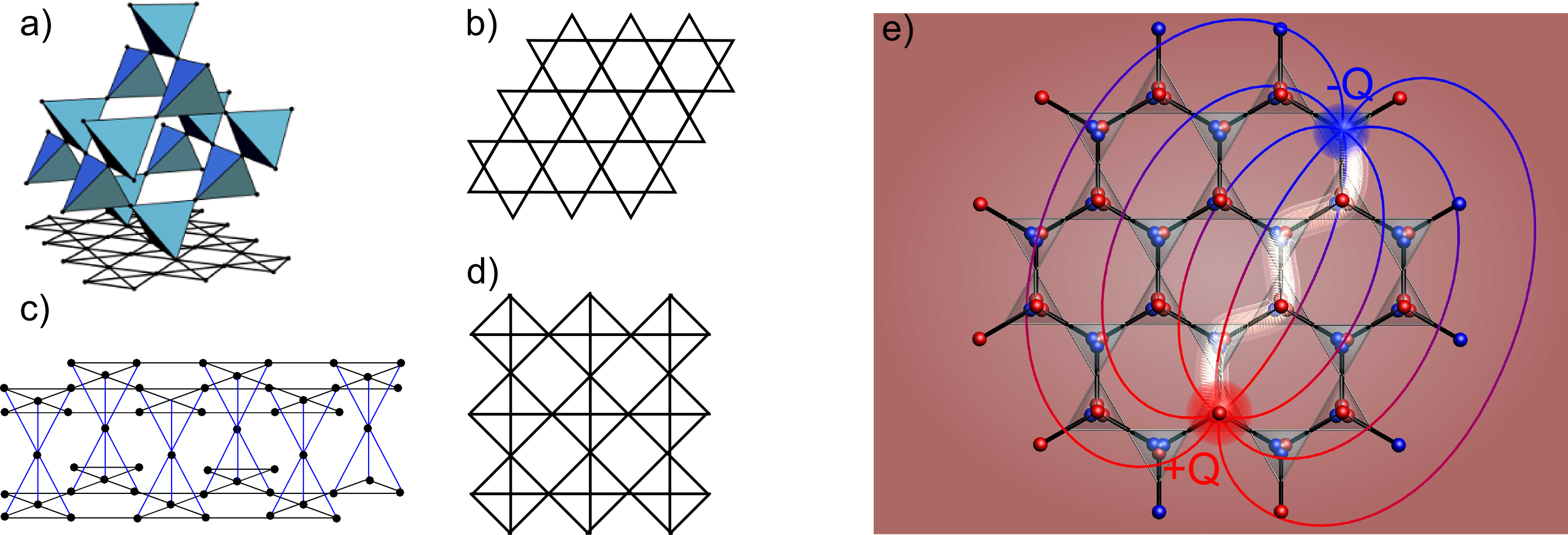}
\caption{Left: Various kinds of corner-sharing lattices with highly degenerate ground states:
a) pyrochlore; b) kagome; c) the pyrochlore slab, as occurs
in the material SrCr$_{9p}$Ga$_{12-9p}$O$_{19}$ (SCGO); d) checkerboard.
e) Cartoon of local violation of Gauss law leading to emergent gauge
charges (magnetic monopoles), which come along with an observable
"Dirac string" connecting them~\cite{castelnovo2008magnetic}. Here,
the magnetic moment of a spin is respresented by a 'dumbell' of
opposite magnetic charges. The net charge of tetrahedra obeying the
ice rules (two spins pointing in and two out) vanishes. However,
flipping the spins along the `Dirac string' generates a pair of
tetrahedra with opposite net magnetic charge. These form freely mobile
quasiparticles known as magnetic monopoles on account of their mutual
magnetic Coulomb interaction.}
\label{fig1}
\end{figure}

On the particular class of lattices consisting of corner
sharing frustrated units containing $q$ spins (where $q\geq3$,
see Fig.~\ref{fig1}),
an n-component spin model can be rewritten in a particularly
intuitive form:
\begin{align}
H = J\sum_{\la i,j\ra}\vec{S}_i\cdot\vec{S}_j
  = \frac{J}{2}\sum_{\alpha} \vec{L}_{\alpha}^2 + \text{const.},
\label{eq:cornHam}
\end{align}
where $\vec{L}_{\alpha} = \sum_{i\in\XBox_{\alpha}}\vec{S}_i$
is the total spin on a single frustrated unit (e.g., a tetrahedron
in the case of a pyrochlore lattice, or a triangle for a kagome
lattice, Fig.~\ref{fig1}), and the variables are n-component vectors.
The local constraints characterizing the ground state
configurations are then explicitly:
\begin{align}
  \vec{L}_{\alpha} = 0, ~\forall \alpha.
\label{eq:gsConst}
\end{align}
This set of equations can be turned into a conservation law
of charges living on the dual lattice (e.g., in the case of
pyrochlore, on the diamond lattice). For that, the original
spin variables are mapped onto fields $\vec{B}^i$ reading on the links of the dual lattice~\cite{isakov2004dipolar},
in such a way that each component $i$ of $\vec{L}_{\alpha}$
is described by a Gauss law:
$\left.\vec{\nabla}\cdot\vec{B}^i\rb_{\XBox_{\alpha}} = \eta(\XBox_{\alpha})L_{\alpha}^i$.
The factor $\eta(\XBox_{\alpha})$ is a sublattice dependent
staggering factor: it equals $+1$ ($-1$) if the frustrated unit
$\XBox_{\alpha}$ occupies the A (B) sublattice. It is therefore
implicitly assumed that the frustrated units occupy a bipartite
lattice. In fact, only in this case it is known how to define
the map to the $\vec{B}$ fields~\cite{henley2010coulomb}
giving the Gauss law required above.

The constraints, Eq.~\ref{eq:gsConst}, can then be resolved by
an emergent gauge field $\vec{B}^i = \vec{\nabla}\times\vec{A}^i$.
Furthermore, the spin correlations are recovered by the $\vec{B}^i$
fields if one imposes a magnetostatic action for each of them as the simplest ansatz for a coarse-grained theory:
\begin{align}
\mathcal{S} = \frac{K}{2}\int d^3x\lnb\vec{B}\rnb^2 .
\label{eq:magnAct}
\end{align}
This is the prominent {\it Coulomb phase}~\cite{isakov2004dipolar,henley2010coulomb} action. It has the form of an emergent Maxwell magnetostatics advertised
above, stabilised by fluctuations between classical ground
states. Here, the gauge theory nature of Maxwell's equations comes
about for purely energetic reasons -- it follows from imposing the
ground-state constraint. Adding quantum fluctuations to such a system
is one way to generate the conjugate electric potential and to produce
Maxwell electrodynamics also hosting, e.g., emergent
photons.\cite{moessner2003three,hermele2004pyrochlore}.

\begin{figure}
    \centering
	\includegraphics[width=1.2\dimexpr11\textwidth/16\relax]{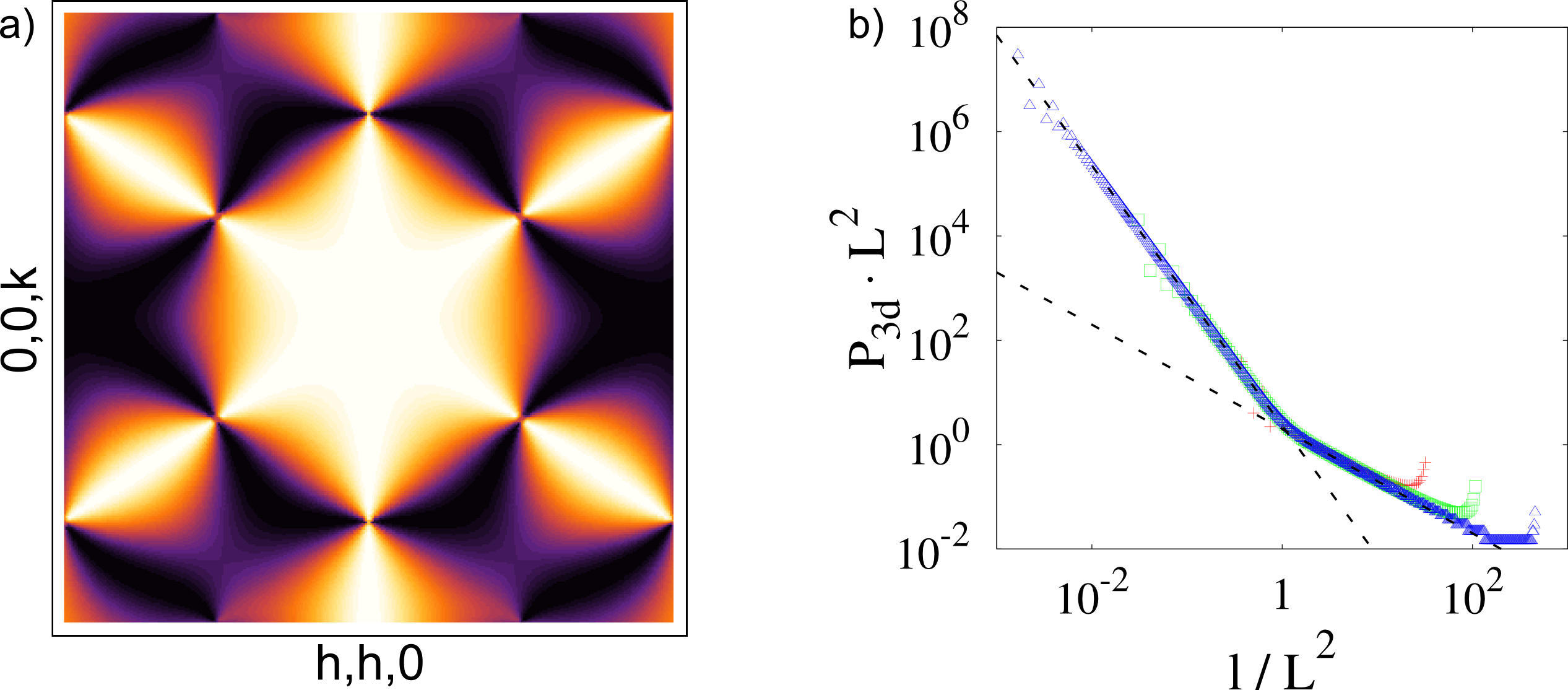}
\caption{a) Large-N prediction for the zero temperature structure factor
on the pyrochlore lattice in the [hhk] plane in Fourier space.
The signature of the emergent magnetostatics in the zero temperature structure
factor of systems presenting a Coulomb phase are the pinch points.
b) Probability, $P_{3D}$, 
distribution of Dirac string lengths, $l$ in spin ice systems of various linear sizes $L$ with periodic boundary
conditions. The population
of strings at large length corresponds to strings with a radius of gyration larger than $L$.}
\label{pinch+dirac}
\end{figure}

Demanding that, the field $\vec{B}$ be divergenceless,
implies for its Fourier modes the condition $\vec{q}\cdot\vec{B}=0$,
and this together with the quadratic action, gives immediately
that pair correlations of the Fourier modes must have the form:
\begin{align}
\la B^{\alpha}_{\vec{q}} B^{\beta}_{\vec{k}}\ra
= \frac{1}{K}\lnb\delta_{\alpha\beta} - \frac{q_{\alpha} q_{\beta}}{q^2}\rnb\delta_{\vec{q},-\vec{k}}
\end{align}
which in real space translates into a dipolar form. This power
law decay translates into characteristic features in the $T=0$
structure factor of the original spin model, that
hallmark all the systems presenting a Coulomb phase, the
so-called {\it pinch-points} (Fig.~\ref{pinch+dirac}, left panel).

Due to Gauss' law, the gauge charges themselves are
$\mathcal{Q}_{\alpha}=\eta(\XBox_{\alpha})L_{\alpha}$, and these
correspond to excitations on the original spin system,
which can be induced thermally or, in a quenched manner, by
diluting the original spin system as described below.

Tracing the magnetostatic action, Eq.~\ref{eq:magnAct},
with the sole constraint of fixing two charges $\mathcal{Q}_{1,2}$
at positions $\vec{r}_{1,2}$ one obtains at zero temperature
(when the $\vec{B}$ field in all other places is divergenceless)
that these charges are subject to an action for two Coulomb charges:
\begin{align}
\mathcal{S}_{\text{int}}(\vec{r}_1,\vec{r}_2) \sim \frac{\mathcal{Q}_1\mathcal{Q}_2}{4\pi\lb\vec{r}_1-\vec{r}_2\rb}
\label{eq:coulInt}
\end{align}

Furthermore, a simple consequence of the staggering factor in the
gauge charges definition is that the following condition must hold
\begin{align}
\sum_{\alpha}\mathcal{Q}_{\alpha} = 0,
\label{eq:chNeut}
\end{align}
akin to global charge neutrality of the universe as a whole.

Up to now we had in mind systems with discrete Ising or continuous
Heisenberg $O(3)$ vector spins leading to one or three flavours for
the gauge charges; the latter are therefore continuous variables.
These charges represent gapless excitations on the original spin system.

The Ising case leads to gapped, discrete
charges. This occurs in nature as a result of a local
easy-axis anisotropy combined to ferromagnetism~\cite{harris1997geometrical} as occurs
in the spin ice materials Ho$_2$Ti$_2$O$_7$, Dy$_2$Ti$_2$O$_7$ \cite{bramwell2001spin}.
In these materials, the spin variables can be described as
$\vec{S}_{i\alpha}=\sigma_{i\alpha}\hat{e}_{\alpha}$, with
$\sigma_{i\alpha}$ being Ising variables, and
$\hat{e}_{\alpha}$ describing the local easy axis, defined
along the bonds of the diamond lattice, dual to the original
pyrochlore lattice where spins reside. The Hamiltonian
has a nearest neighbor contribution, as well as a
long-ranged, dipolar one~\cite{denHertog2000dipolar}
(borrowing notation from Ref.~\cite{isakov2005why}):
\begin{align}
H = \sum_{\text{pairs}}\sigma_{i\alpha}\lsqb J\mathcal{J}_{i\alpha,j\beta}+Da^3\mathcal{D}_{i\alpha,j\beta}\rsqb\sigma_{j\beta} ,
\label{eq:spIceHam} \\
\mathcal{D}_{i\alpha,j\beta} = \frac{\hat{e}_{\alpha}\cdot\hat{e}_{\beta}}{\lb\vec{r}_{i\alpha,j\beta}\rb^3}
- \frac{3 \lnb\hat{e}_{\alpha}\cdot\vec{r}_{i\alpha,j\beta}\rnb \lnb\hat{e}_{\beta}\cdot\vec{r}_{i\alpha,j\beta}\rnb}{\lb\vec{r}_{i\alpha,j\beta}\rb^5}
\label{eq:dipMat}
\end{align}
with $a$ the nearest neighbor distance, $\mathcal{J}_{i\alpha,j\beta}$
the pyrochlore adjacency matrix, and $\mathcal{D}_{i\alpha,j\beta}$
the interaction matrix of the magnetostatic dipolar interactions.

The emergent gauge charges on these systems have a fascinating physics of
their own, and we devote the next two subsections to them.

\subsection{Magnetic monopoles}

The ground state local constraints on spin ice systems receive
usually the special name of {\it ice rules}, coming from a
natural mapping between these systems, and common water ice,
where the ice rules were first formulated. Local violations of
the ice rules occur by single spin flips which leads to
gapped {\it magnetic monopoles}~\cite{castelnovo2008magnetic}.
These are emergent gauge charges, and on account of the discussion
in previous section also experience a Coulomb entropic force.

Only a finite amount of free energy is necessary to separate monopoles
infinitely, and therefore these are {\it deconfined}~\cite{castelnovo2008magnetic,castelnovo2012spin}
objects. This constitutes a ``fractionalization'' of the original
magnetic dipole moment into magnetic charges.
They can be separated by further spin flips,
thereby leaving behind an observable {"Dirac string"}
~\cite{jaubert2008three,jaubert2009the,morris2009dirac} see Fig. \ref{fig1}.
Indeed, an easy way mathematically to see the genesis of magnetic
charges is to observe that inverting a string of dipoles sets up
a potential equivalent to that of two equal and opposite charges,
$\pm Q$, at its ends. In a continuum approximation, the dipole moment
density along the string corresponds to one atomic magnetic
moment, $\mu$ per bond of the dual diamond lattice, which has
length $a$; since a string is flipped, rather than considered
in isolation, there is an additional factor of 2: $Q=2\mu/a$.

The Dirac strings themselves are loosely defined objects, as a given
configurations does not define them uniquely. Nevertheless it is
possible to define them stochastically, and study the statistics of
their lengths $l$, which present a broad distribution, with two
power-law regimes: short loops with a probability distribution
scaling as $\sim L^3l^{-2.50(1)}$ and long loops with a scaling of
$\sim l^{-0.98(3)}$ (Fig.~\ref{pinch+dirac})~\cite{jaubert2011analysis}.
The effective low temperature system can therefore be seen as
a soup of magnetic monopoles wandering around, with the
``spaghetti'' of Dirac strings fluctuating between them.
The power-law form of their length distribution reflects
their thermodynamic tensionlessness, and underpins the
deconfinement of the magnetic monopoles.

A perhaps illuminating approach to understand the origin of
the Coulomb interaction from the dipolar one is obtained
through the simplified ``dumbell model''~\cite{castelnovo2008magnetic,henley2010coulomb}
Figs.~\ref{fig1},\ref{fig:ghost}. Within this model, the original spins having magnetic
moment $\mu$, are replaced by a dipole of charges $\pm Q/2$
separated by the diamond lattice bond spacing $a$, so that
the effective dipole moment of this dumbell coincides with
the original spin magnetic moment, which fixes the 
charges to be $Q=2\mu/a$, the value derived above.

The dumbell model is useful, since monopole charges are directly
given by the sum of dumbell charges on a tetrahedron. The original
dipolar interactions between spins are translated into effective
Coulomb interactions between these monopoles. This interaction is
therefore of an energetic origin, contrary to its entropic
counterpart originating from an averaging over the many
degenerate configurations compatible with defects placed at
fixed positions.

\subsection{Intrinsic versus emergent gauge charge}

The dumbell model has led us to the important distinction between
two contributions to the monopoles' interactions. In fact these
two contributions correspond to two kinds of charges in our system.

The emergent gauge charges exist due to the cooperative behavior
of the whole system, which on account of the non-trivial ground
state correlations, leads to entropically interacting magnetic
monopoles.

On the other hand, the intrinsic gauge charges interact on
energetic grounds alone, as a result of the original dipolar
interactions for the magnetic dipoles constituting the system.
The ground state constraints let these original dipolar
interactions, decaying as $1/r_{ij}^3$, be described by effective
Coulomb interactions between the intrinsic gauge charges,
thus with a slower decay, decaying as $1/r_{ij}$.

\section{Disorder in a spin liquid}
\label{sec:disSL}

The most unavoidable way for the system to disobey Gauss' law
 is on account of thermal fluctuations. Nonetheless,
the original microscopic model provides another manner for
violating these constraints, robust, down to the zero temperature limit. This is
achieved on diluting the original system.

In contrast to our conventional Maxwell vacuum, consideration of
defects on the underlying space in a condensed matter
system is quite natural, since the materials, which we
are ultimately interested in describing by our simplified
models, very often present unavoidable, or deliberately introduced, non-magnetic impurities.

Two important cases must be distinguished here, according to
whether we dilute a lattice presenting discrete or continuous spins.

\begin{figure}
    \centering
	\includegraphics[width=1.2\dimexpr11\textwidth/16\relax]{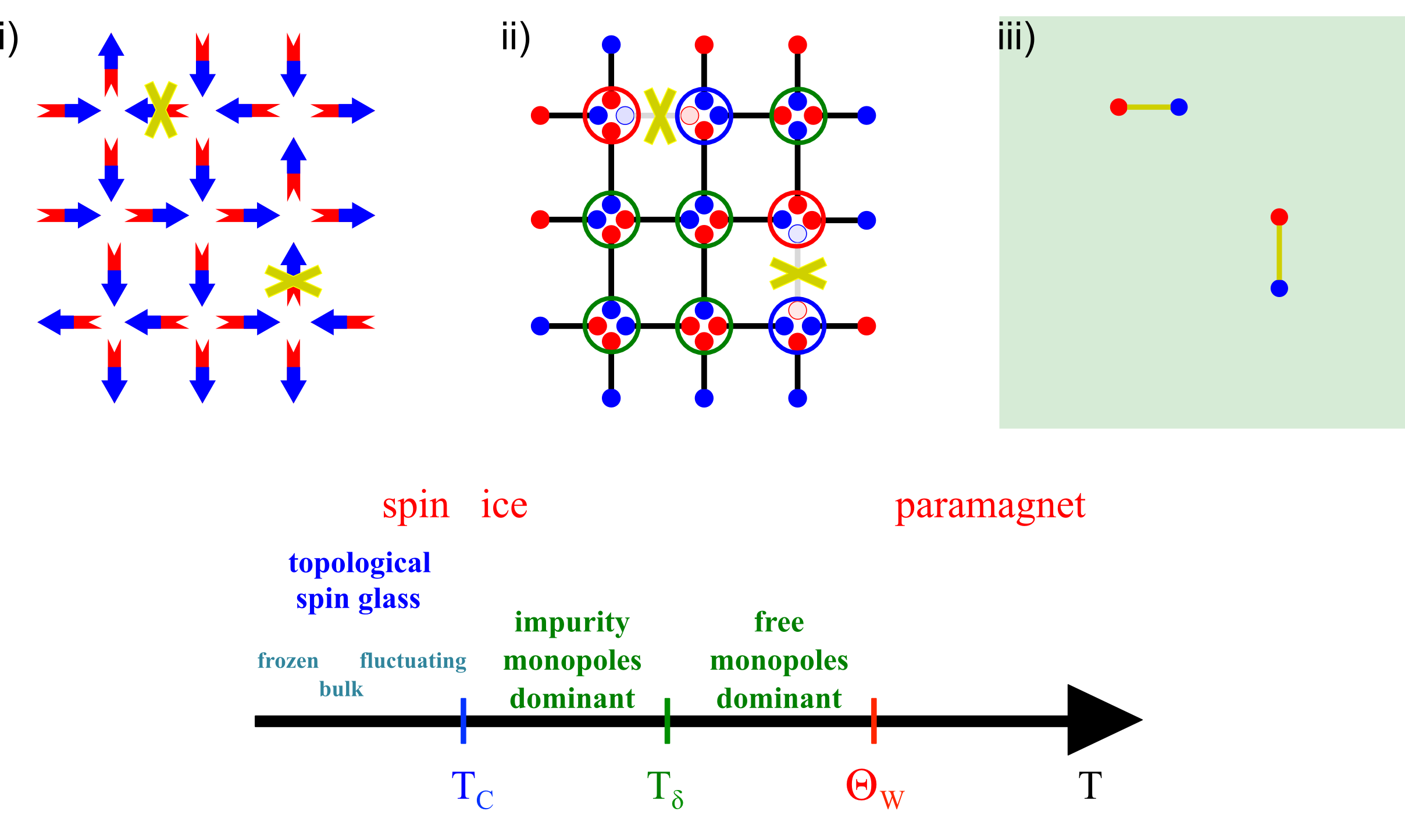}
\caption{Top: Steps
onto effective description of quenched defects in spin ice~\cite{sen2015topological}:
i) missing spins due to random dilution; ii) in the dumbell picture
the missing spins yield sites on the dual lattice with odd coordination,
which therefore cannot be charge neutral for discrete spins; iii) at low
temperatures only these are the relevant degrees of freedom once the
fluctuating background of charge-neutral tetrahedra (denoted
in green) has been integrated out. These correspond to `ghost spins'
at random locations given by the locations of the missing spins!
Bottom: Phase diagram proposed for diluted spin ice.}
\label{fig:ghost}
\end{figure}

\subsection{Topological spin glass}

The ground states of spin-ice systems are characterized by the
ice rules which can only be violated by a finite energy gap, $\Delta$,
e.g., by flipping a single spin. This corresponds to creation of two
nearest-neighbour monopoles with opposite charges. Interestingly,
if instead of flipping a spin, a single spin in a configuration
satisfying the ice rules is removed, the same effect is produced
in the alternative monopole picture--two nearest-neighbor monopoles with
opposite (half-)charges $Q/2$ are created. Still an important difference
here applies, namely, the ice-rules where the monopoles were created
cannot be restored by further spin flips, as in the disorder-free
case, since each of these frustrated units contains now an odd
number of discrete spins and with this, $L=0$ is impossible.
At low temperatures, when thermally activated monopoles become
exponentially suppressed, only the quenched monopoles due to disorder
are relevant. A quenched pair of nearest-neighbor oppositely charged monopoles
corresponds therefore to an emerging {\it ghost-spin}~\cite{sen2015topological},
which becomes the relevant degree of freedom, while all the original
spins can be integrated out in this limit, Fig.~\ref{fig:ghost}.

The randomly placed ghost spins mediate through the correlated
background dipolar interactions, which have two contributions as
 for the usual monopoles, one entropic, due to the fluctuations
among the degenerate ground states, and one energetic, due to long
ranged dipolar interaction present on spin ice compounds. The latter
survives even down to the zero temperature limit, while the entropic
contribution vanishes linearly with $T$.

This low temperature description for spin-ice in terms of ghost spins
turns out to present a spin-glass phase transition at non-zero
temperatures (dipolar spin glass), where the ghost spins freeze,
while the remaining bulk is still able to fluctuate~\cite{sen2015topological}. The
freezing transition is found to depend linearly on the dilution,
$T_c(x)\propto x$. This can be simply understood in terms of the
typical energy scale of the dipolar interaction, scaling as
$H_{dip}\sim 1/r_{typ}^3$, while $r_{typ}^3\sim 1/x$.

\subsection{Orphan spins}

The consideration of dilution in a Coulomb phase system with
vector charges has a long history~\cite{moessner_ramirez} with the
first systematic experimental exploration being made by Schiffer
and Daruka on the material SCGO~\cite{schiffer1997two}, even
before the Coulomb phase was properly identified. They proposed
a two-population model on explaining the observed uniform
magnetic susceptibility a population of correlated spins
coexisting with uncorrelated ones, leading to a Curie
tail $T^{-1}$ on the susceptibility, the so-called {\it orphans}.

The single-unit approximation proposed~\cite{moessner1999magnetic}
shortly after these experimental findings, led to the
conclusion that dilution in these systems can lead to
orphans on those frustrated units where all but one spin
is left alone ($L=0$ is impossible only for $q=1$). These are
still correlated with the rest of the system through
the remaining frustrated unit to which it belongs.

\begin{figure}
    \centering
	\includegraphics[width=1.2\dimexpr11\textwidth/16\relax]{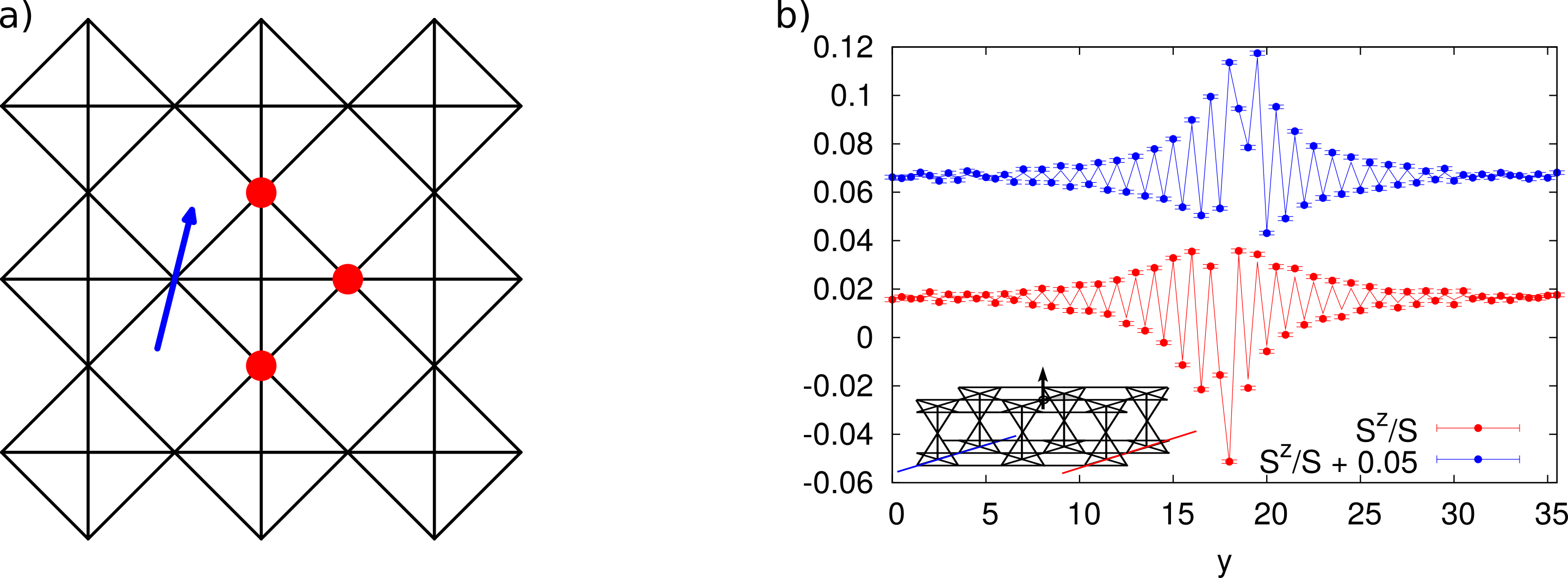}
\caption{a) Orphan spin, emerging from diluting a lattice with corner
sharing structure~\cite{rehn2015random} such that a spin has no
interaction partner in one of the simplices that share it.
b) The texture originating around the orphan placed on the
magnetic lattice of SCGO~\cite{sen2011fractional}: spin density along the coloured
lines in the presence of an orphan spin located as indicated (the blue curve offset
for better visibility). Obtained from low-tempeature Monte Carlo simulations.}
\label{fig:orp}
\end{figure}

This correlation of the orphan with its spin-liquid ``bath'' has
important consequences as~\cite{sen2011fractional,sen2012vacancy} a texture
emerges around each orphan, and partially screens it,
effectively reproducing an orphan+texture fractionalized object,
with a fractional moment of $1/2$ of the original
spin moments (Fig. \ref{fig:orp}).

In the magnetostatic picture, this texture is a natural
consequence of Gauss law: due to the defect charge (orphan),
$\mathcal{Q}$, an emergent magnetic field originates around it.
In $d$ dimensions, since $\oint \vec{B}\cdot d\vec{S} \propto \mathcal{Q}$,
the field decays as $1/r^{d-1}$ with distance, and this is
precisely the scaling found for the texture at $T=0$.
Despite this slow decay, the texture forms a total moment
conspiring to cancel only half of the orphan moment, and
this is due to the oscillations coming from the staggered
definition of the local fields, see Fig.~\ref{fig:orp}.

\subsection{Interactions between orphan spins}

The orphans are disorder nucleated gauge charges and as such,
are subject to mutual interactions mediated by the background
Coulomb phase in which they are embedded. These are obtained
in the the same way as considered before, by integrating out
the remaining spins, which thus serve as ``bath'' to the
orphans, leaving an effective description for the orphans
alone, which, not surprisingly, interact as vector charges:
\begin{align}
\beta J_{eff}(r_{ij}) \eta_i\vec{L}_i \cdot \eta_j\vec{L}_j
\end{align}
with the sublattice dependent staggering factors $\eta_i$
explicitly shown. This staggering is removable
by a trick due to Mattis~\cite{mattis1976solvable}: define
new spin variables on one of the two sublattices to
be equal to the reverse of the old spin variables. This
Mattis transformation removes the sublattice dependence at
the cost that certain observables (such as the magnetization)
obtain a modified meaning (staggered magnetization).

This leads at zero temperature to the effective random
Coulomb antiferromagnet model describing the system of
orphans, which has been studied in detail on Ref.~\cite{rehn2015random}.

The limit to zero temperature is necessary as only on this
limit the interactions become truly long-ranged, as they
are at finite temperatures thermally screened. More precisely
one encounters in this limit orphans interacting with:
\begin{align}
\lim_{T\rightarrow 0}\beta J_{eff} = A J_{ij}
\end{align}
with $J_{ij}$ the Coulomb potential between orphans $i$ and $j$,
and the coupling strength $A$ being fixed by
this limit and the original microscopic model parameters.

\subsection{Random Coulomb magnets}

The new models describing effective charges, induced by dilution
on a Coulomb phase with continuous charges, are thus of the form (at $T=0$):
\begin{align}
H = \frac{1}{2}\sum_{i,j} J_{ij} \vec{L}_i\cdot\vec{L}_j,
\label{eq:rcm}
\end{align}
with the properties
\begin{itemize}
\item $J_{ij}>0, ~\forall i,j$ (after a Mattis transformation).
\item $J_{ij}\sim 1/\lb\vec{r}_i-\vec{r}_j\rb$ in three, or
	 $J_{ij}\sim\log{\lb\vec{r}_i-\vec{r}_j\rb}$ in two dimensions.
\item the spin positions, $\vec{r}_i$, are quenched random variables.
\end{itemize}
These models we call the {\it random Coulomb magnets} (RCM)
a new type of frustrated disordered magnetic model which naturally
arises in the context of diluted Coulomb phases.

The interaction matrices on these effective models belong to
the class of Euclidean random matrices, which are generally
characterized by the dependence of $J_{ij}$ on the Euclidean distance
$J_{ij}=J(|\vec{r}_{ij}|)$, while the randomness lies alone on
the independently distributed random variables $\vec{r}_i$~\cite{mezard1999spectra,goetschy2013euclidean}.
The properties of this ensemble are comparatively not as
intensively studied in the literature of spin-glass models
as the commonly considered random matrices there, which usually
involve the important different assumption of independence between
the different matrix elements. Also, in Eq. (3.3), all interactions
are antiferromagnetic -- we have a random Coulomb antiferromagnet,
unlike the usual spin glass Hamiltonian with random signs of the
magnetic interaction.

These models resemble plasma models, with the fluctuating degrees
of freedom being instead of the charged particle positions, the ``signs'' of the 
charges themselves, which can then be allowed to take on values given by
 $O(N)$
symmetry. The case $N=3$ corresponds thus to the original microscopic
model giving birth to these models. The same model in two dimensions
for the case $N=1$ has been found as the effective description of
disorder induced defects on a completely different microscopic model
many years ago by Villain~\cite{villain1977two}.

The global charge neutrality constraint on the microscopic theory,
Eq.~\ref{eq:chNeut}, translates into (after the Mattis transformation):
\begin{align}
\sum_{i} \vec{L}_i = 0.
\label{eq:orpNeut}
\end{align}
This is quite a natural constraint, which even without explicit imposition in
the original microscopic model, alone on energetic grounds
of the long ranged Coulomb model described by Eq.~\ref{eq:rcm} would be very nearly satisfied.

The RCM's have been studied in some detail in Ref.~\cite{rehn2015random},
both for the cases $N=3$ and $N\rightarrow\infty$, the latter
also serving as a semi-analytical proofing of the (necessarily)
purely numerical results on $N=3$.

With the aim of understanding the general phase diagram of
these models, Ref.~\cite{rehn2015random} allows the coupling
strength $A$ to vary, while the density of orphans $x$ is
fixed but kept small (the behavior at some high enough
values of $x$ leads to trivially ordered, striped or Neel
phases), in the expectation of possibly detecting a
spin-glass ordering as one increases $A$.
This quest is inspired by a proposal already made by Schiffer
and Daruka in their original work on the orphans, that these might
actually be degrees of freedom undergoing freezing known to 
occur on SCGO at low temperatures, while the bulk of spins could
well keep their own dynamics, and not necessarily freeze at the same
temperature as the orphans. These experimental observations
on SCGO and similar highly frustrated magnetic materials,
constitute a very old puzzle waiting for explanation already for
several decades~\cite{obradors1988magnetic,ramirez1990strong,laforge2013quasispin}.
Furthermore, the idea that a spin glass phase might arise out of
emergent degrees of freedom originating on an insulating matrix
provided by a highly frustrated magnetic system goes much further
back in time, as argued by Villain on Ref.~\cite{villain1979insulating}.

The simulations in Ref.~\cite{rehn2015random} found a broad
paramagnetic regime for the random Coulomb magnets and up to
very large values of $A$, no spin glass phase transition for
a finite value of $A$ was found  both in two and three
dimensions in systems with $N=3$. Furthermore in two dimensions,
the system with $N\rightarrow\infty$ indicated the critical
glassy behavior occurring for $A\rightarrow\infty$, although
no conclusive statements have so far been obtained for three dimensions.

Although numerics in these models can not explain orphan
freezing, one particularly interesting property on the
two-dimensional case is that at very low orphan densities
$x$, a global
scaling transformation, $r_{ij}\rightarrow\kappa r_{ij}$
of the orphan distances is innocuous on account of the
logarithmic form of the interactions, and the global
charge neutrality,~\ref{eq:orpNeut}, thereby leaving the
partition function unchanged up to an irrelevant constant factor:
\begin{align}
Z' =  e^{A\log(\kappa)M/2}Z ,
\end{align}
with $M$ the total number of orphans. This implies that in
the limit of low densities $x$, any critical coupling $A_c$
will be independent of $x$: the partition function itself is a scaling function.

The analogy of these models to plasma physics can be pushed
further and a general screening theory in same spirit of
Debye-H\"uckel theory is developed in Ref.~\cite{rehn2015random},
with the result that due to spin fluctuations alone
(note that the spins are quenched on random lattice positions)
a thermal screening length originates, scaling in a similar
way to the Debye-H\"uckel screening length as $\xi\sim1/\sqrt{A}$.
The necessity for screening in these models can be seen directly
in a high temperature expansion (HTE), as on account of the long
rangedness of the interactions, the series of the HTE must be
resummed on each term to remove a divergence.

In summary, RCM's present a very rich physics, with connections
to several areas of physics, such as the physics of plasmas
and Debye-H\"uckel screening, or the physics of
spin-glass models and the relatively poorly explored ensemble
of Euclidean random matrices.

\section{Summary and outlook}
\label{sec:concl}

In this article, we have showcased how Maxwellian physics arises in
particular, geometrically frustrated, condensed matter systems. We
have given some examples of how it behaves like conventional
electromagnetism, and how it can appear to be richer, such as in the
case of the emergence of freely mobile magnetic monopoles from the
fractionalisation of spin moments, which in addition to their magnetic
charge carry an emergent Coulomb gauge charge.

We have also discussed how new degrees of freedom can emerge when
disorder is added to the system, and how the Coulomb spin liquid which
hosts them mediates interactions between these. This leads to the
appearance of unusual ghost spin degrees of freedom, which encode the
missing spins much in the same way as holes in a band insulator
represent missing electrons. These ghost spins themselves undergo a
freezing transition. We have also introduced the rich physics of
random Coulomb magnets, which is only beginning to be explored.

The study of quenched disorder in the lattice is a particularly
attractive feature of this class of models, as it has a natural
interpretation in terms of defects in the `ether' underpinning the
Coulomb phase.

Varying the lattice then allows for the emergence of different types
of disorder physics. For instance, in all the cases currently known,
where the map from spin variables to an emerging field is described
by a magnetostatic action, one important requirement is that the
frustrated units occupy a {\it bipartite lattice}~\cite{henley2010coulomb}.
This has an important consequence for the gauge-charges, as their
lattice scale dependence is completely removable, and their
long wave-length description is purely the one of emerging
Coulomb charges on the continuum.

In this sense, it is very interesting that at least one system has
been recently recognized where the emerging gauge-charges interact
as emerging Coulomb charges~\cite{rehn2015classical} but the
frustrated units occupy a non-bipartite lattice. Here, the
fractionalisation of the spin moments is into orphan spins
carrying an odd-denominator fraction, 1/3, of the microscopic magnetic
moments, the first time such fractionalisation has been observed in a
classical spin system.

We hope that the study of Coulomb spin liquids will continue to
produce further such surprises, and continue to allow condensed matter
physics to provide novel aspects to add to the venerable field of
Maxwell electromagnetism.

%==============================================================================%
\contributions{Both authors contributed extensively to this manuscript.}

\conflict{The authors declare they have no competing interests.}

\funding{This work was supported by the Deutsche Forschungsgemeinschaft
 under grant SFB 1143.}

\ack{We thank our collaborators on the various topics covered here:
Alex Andreanov, Claudio	Castelnovo, John Chalker, Kedar Damle, Karol Gregor, 
Masud Haque, Peter Holdsworth, Sergei Isakov, Ludovic Jaubert, Anto Scardicchio, 
Arnab Sen, Shivaji Sondhi and Peter Young.}
%==============================================================================%

\bibliographystyle{unsrt}
\bibliography{ref_maxwell}

%==============================================================================%
\end{document}